%% file: main.tex
\pdfoutput=1
\documentclass[sigconf]{acmart}

\AtBeginDocument{%
  }

\usepackage{enumitem}
\usepackage{threeparttable}
\usepackage{multirow}
\usepackage{bm}
\usepackage{subfigure}
\usepackage{textcomp}%
\usepackage{changepage}
\usepackage{xcolor}

\usepackage{algpseudocode}
\usepackage{algorithmicx,algorithm}
\usepackage{mathtools}

\usepackage[group-separator={,}]{siunitx}

\usepackage[justification=justified,skip=3pt]{caption}

\copyrightyear{2025}
\acmYear{2025}
\setcopyright{cc}
\setcctype{by}
\acmConference[XXX]{XXX}{XXX}{XXX}
\acmBooktitle{XXXXX}
\acmDOI{XXXXX}
\acmISBN{XXXXX}

\begin{document}

\title{From Generation to Consumption: Personalized List Value Estimation for Re-ranking}

\author{Kaike Zhang}
\affiliation{%
  \institution{Kuaishou Technology}
  \institution{University of Chinese Academy}
  \country{of Sciences, Beijing, China}
}
\email{kaikezhang99@gmail.com}

\author{Xiaobei Wang}
\affiliation{%
  \institution{Kuaishou Technology}
  \country{Beijing, China}
}
\email{wangxiaobei03@kuaishou.com}
\authornotemark[1]

\author{Xiaoyu Yang}
\affiliation{%
  \institution{Kuaishou Technology}
  \country{Beijing, China}
}
\email{yangxiaoyu@kuaishou.com}

\author{Shuchang Liu}
\affiliation{%
  \institution{Kuaishou Technology}
  \country{Beijing, China}
}
\email{liushuchang@kuaishou.com}
\authornote{Corresponding author}

\author{Hailan Yang}
\affiliation{%
  \institution{Kuaishou Technology}
  \country{Beijing, China}
}
\email{yanghailan@kuaishou.com}

\author{Xiang Li}
\affiliation{%
  \institution{Kuaishou Technology}
  \country{Beijing, China}
}
\email{lixiang44@kuaishou.com}

\author{Fei Sun}
\affiliation{%
  \institution{University of Chinese Academy}
  \country{of Sciences, Beijing, China}
}
\email{ofey.sunfei@gmail.com}

\author{Qi Cao}
\affiliation{%
  \institution{University of Chinese Academy}
  \country{of Sciences, Beijing, China}
}
\email{caoqi92seven@gmail.com}

\renewcommand{\shortauthors}{Kaike Zhang et al.}

\begin{abstract}

\input{Section/0-Abstract}
\end{abstract}



\maketitle

\section{INTRODUCTION}
\input{Section/1-Introduction.tex}

\section{RELATED WORK}
\input{Section/2-Related_Work.tex}

\section{PRELIMINARY}
\input{Section/3-Preliminary.tex}

\section{METHOD}
\label{sec:method}

\input{Section/4-Method.tex}

\section{BENCHMARK}
\label{sec:bench}
\input{Section/5-Benchmark.tex}

\section{EXPERIMENTS}
\input{Section/6-Experiment.tex}

\section{CONCLUSION}
\input{Section/7-Conclusion.tex}

\bibliographystyle{ACM-Reference-Format}
\bibliography{ref}


\end{document}

%% file: Section/0-Abstract.tex
Re-ranking plays a pivotal role in recommender systems by optimizing the ordering of recommendation lists, thereby enhancing user satisfaction and platform revenue. Most existing approaches adopt a generator-evaluator framework, where the generator produces multiple candidate lists and the evaluator selects the optimal one by estimating the entire list value. However, such methods often overlook the subjectivity and randomness of user behavior, i.e., users may not fully consume the entire list. This oversight leads to a discrepancy between the estimated \textbf{generation value} and the \textbf{actual consumption value}, finally resulting in suboptimal recommendations. To address this gap, we propose modeling the user’s exit probability to better align the estimated list value with actual user consumption. Based on this insight, we propose a personalized \textbf{C}onsumption-\textbf{A}ware list \textbf{V}alue \textbf{E}stimation framework (\textbf{CAVE}), which formulates the consumption value of a list as the expectation over the values of its sub-lists, weighted by the user’s exit probability at each position.  Specifically, CAVE decomposes the user exit probability into two components: an interest-driven part reflecting user engagement, and a stochastic part capturing random external factors, the latter of which is modeled using the Weibull distribution. By jointly modeling sub-list values and exit probabilities, CAVE provides a more accurate estimate of the list's actual consumption value. To facilitate further research, we construct and publicly release three real-world list-wise benchmarks from the Kuaishou platform, covering varying dataset sizes and user activity distributions. Extensive offline experiments on these benchmarks and two widely used Amazon datasets, as well as online A/B tests on the Kuaishou platform, demonstrate that CAVE consistently outperforms state-of-the-art baselines, underscoring the importance of modeling user exit behavior in re-ranking.

%% file: Section/1-Introduction.tex

Recommender systems have become essential tools for managing the exponential growth of information available online~\cite{covington2016deep, gomez2015netflix,gong2022real}. Typically, recommender systems follow multiple stages~\cite{qin2022rankflow,wang2011cascade,zheng2024full}, including  matching~\cite{koren2009matrix, he2020lightgcn, zhang2024understanding}, ranking~\cite{yu2019multi, cheng2016wide, zhou2019deep}, and re-ranking~\cite{zhuang2018mutual,ai2018listwise,liu2023gfn4list}. Among these, re-ranking plays a crucial role as it directly determines the final recommendation results shown to users, thereby significantly influencing user satisfaction and overall platform revenue.

\begin{figure}
\centering
\includegraphics[width=\linewidth]{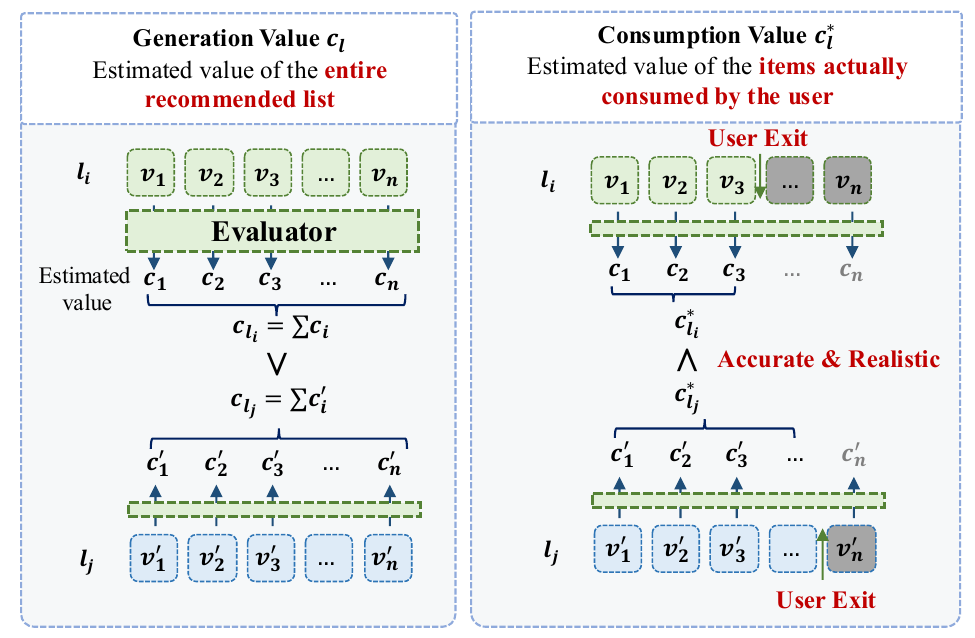}
\caption{From generation value estimation to consumption value estimation. Left: Generation value---the estimated value of the entire generated list. Right: Consumption value---the estimated value of items actually consumed by the user.}
\label{fig:motivation}
\end{figure}

Existing re-ranking methods can be broadly categorized into single-stage, which normally stated as Generator-only~(G-only)~\cite{zhuang2018mutual,ai2018listwise,pei2019prm,gong2022edgererank,liu2023gfn4list} and two-stage approaches~\cite{shi2023pier,xi2024utility,lin2024dcdr,ren2024nar}. G-only methods directly generate recommendation lists by scoring items and arranging them greedily~\cite{gong2022edgererank,liu2023gfn4list}. However, this approach often ignores interdependencies among items, resulting in suboptimal rankings~\cite{shi2023pier,xi2024utility}. Consequently, two-stage pipelines have increasingly gained attention. Two-stage methods typically follow a generator-evaluator~(G-E) framework~\cite{shi2023pier,xi2024utility,lin2024dcdr,ren2024nar}. Initially, the generator produces multiple candidate recommendation lists, after which the evaluator selects the optimal list based on an estimated list-wise value. This separation of list generation and evaluation has shown significant improvements in ranking quality. In such a framework, since the evaluator directly determines the final list selection, it is crucial to ensure the accuracy of the evaluator’s estimation of list value. 

However, current evaluators typically estimate the value of the \textbf{entire generated list}, without considering how user behavior influences consumption in the recommendation loop. In practice, \textbf{users may not fully consume the entire list}, as they may exit partway through the list. This means the estimated list-wise value does not always reflect the actual consumption value experienced by the user, leading to inaccurate evaluations and potentially suboptimal re-ranking decisions.
As illustrated in the left part of Figure~\ref{fig:motivation}, suppose that List $l_i$ has a higher estimated total value than List $l_{j}$. However, as shown on the right, when users exit early, List $l_{j}$ can yield a higher actual consumption value. Such discrepancies between estimation and real user behavior often result in suboptimal re-ranking outcomes.

To address this gap, we propose estimating the consumption value of the list, aiming to more accurately reflect the value that the list brings to users. Specifically, we model the user’s exit probability during the interaction process, such that the consumption value of a list can be viewed as the expectation over sub-list values, weighted by the user’s probability of exiting at each position. By incorporating user exit probabilities, we aim to bridge the gap between traditional list generation value and actual consumption value.

Building on this insight, we introduce a personalized \textbf{C}onsumption-\textbf{A}ware list \textbf{V}alue \textbf{E}stimation framework, named \textbf{CAVE}. CAVE consists of two core components: a sub-list value estimation module and an exit probability estimation module. The former learns to estimate the value of each sub-list from user feedback, while the latter predicts the likelihood of user exit at each position in the list. To model user exit behavior more precisely, we decompose the exit probability into two distinct components: an \textbf{interest-driven exit probability} that reflects user engagement with the content, and a \textbf{stochastic exit probability} that captures external random factors such as fatigue or distraction. The interest-driven component is learned by minimizing the discrepancy between the predicted consumption value and the observed feedback from users. To capture the stochastic nature of early exits, we incorporate the Weibull distribution\footnote{The Weibull distribution is commonly used to model time-to-event data and is flexible in capturing various hazard rate patterns.}~\cite{weibull1951statistical} to fit the overall distribution of stochastic exit probabilities across users. In this way, CAVE enables more accurate estimation of a list’s expected consumption value, ultimately improving list selection and re-ranking quality.

Additionally, to address the lack of realistic list-wise exposure feedback in existing datasets, we publicly release three large-scale real-world benchmarks collected from the Kuaishou platform, which contain millions of list-wise exposures along with rich user and item features, as well as detailed interaction feedback. 
They differ in both dataset size and user behavior distribution, enabling comprehensive evaluation under various real-world conditions. We conduct comprehensive evaluations on these three newly released datasets, along with two commonly used public datasets. Extensive offline experiments, as well as online A/B tests on the Kuaishou platform, demonstrate that CAVE consistently outperforms state-of-the-art baselines, validating its effectiveness in accurately estimating list consumption value and improving re-ranking performance.

Our main contributions are summarized as follows:
\begin{itemize}[leftmargin=*]
\item We publicly release three real-world benchmark datasets from the Kuaishou platform, containing genuine list-wise user feedback.
\item We introduce a novel objective for list value estimation by explicitly modeling user exit probabilities and incorporating the Weibull distribution to capture stochastic behavior.
\item We propose a unified evaluator framework that achieves superior performance over existing methods, as validated through extensive offline and online experiments.
\end{itemize}

%% file: Section/2-Related_Work.tex
\subsection{Reranking in Recommendation Systems}\label{sec: related_rerank}

Reranking is a critical component in multi-stage recommender systems, aiming to refine initial candidate lists to maximize user satisfaction by modeling mutual influences among items. Early works ~\cite{carbonell1998mmr,joachims2005interpret} recognized that standard learning-to-rank approaches treating items independently fail to capture intra-list correlations essential for optimal exposure. 

To address this, several generator-only approaches ~\cite{zhuang2018mutual,ai2018listwise,pei2019prm,gong2022edgererank,liu2023gfn4list,feng2021revisit,xi2022multi,pei2019personalized} adopt models such as GRU-based DLCM~\cite{ai2018listwise} or Transformer-based PRM~\cite{pei2019prm}, estimating refined scores with listwise features and greedily selecting top items. However, these methods often suffer from inconsistencies between reranked permutations and initial conditioning, leading to suboptimal results.

To overcome such limitations, two-stage reranking methods employing the generator-evaluator~(G-E) framework~\cite{chen2022extr,shi2023pier,xi2024utility,lin2024dcdr,ren2024nar,wang2025nlgr,yang2025comprehensive} have emerged as state-of-the-art solutions. In this paradigm, a generator proposes multiple candidate sequences, and an evaluator assesses their listwise utility to select the best for final exposure. Generative models are preferred over heuristic permutation approaches due to their capacity to efficiently explore the exponentially large permutation space while incorporating user preferences. 
Some recent work also noticed that large language models~\cite{ren2024rlm4rec,gao2024llm,wu2024survey,gao2025llm4rerank,ren2025self,liu2025leveraging} and  reinforcement learning~\cite{feng2021grn,wang2024future,wang2025value,wei2020generator} can enhance the reranking performance by involving textual information, regarding two-stage reranking as actor critic paradigm and decomposing the list-wise value. 

Nevertheless, existing two-stage reranking research has predominantly focused on designing more powerful generators. In contrast, the design of evaluators has received relatively limited attention, often being implemented as simple utility scoring functions or auxiliary discriminators without in-depth exploration of their architectures, and aligning user satisfaction .


%% file: Section/3-Preliminary.tex
\begin{figure*}
\centering
\includegraphics[width=7.0in]{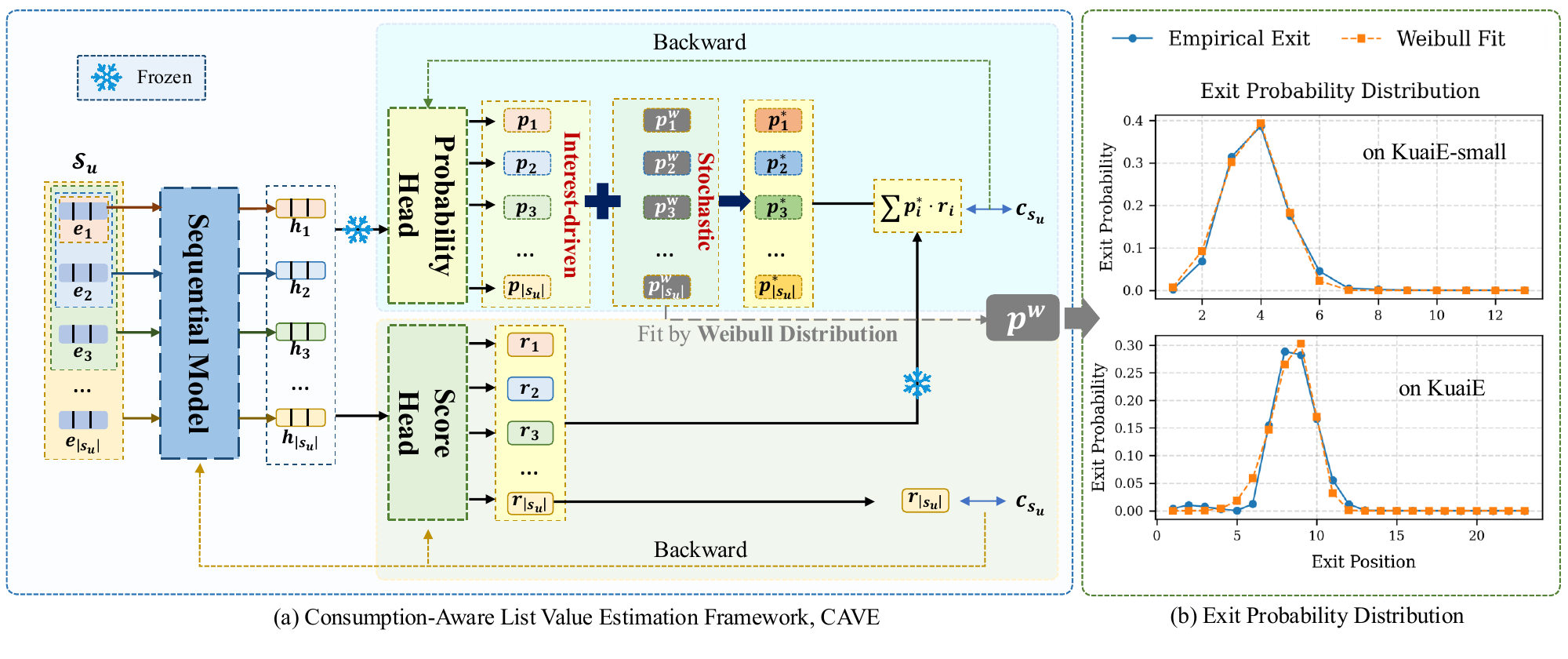}
\caption{Overview of the CAVE Framework. (a) CAVE estimates the actual consumption value by jointly modeling sub-list values and user exit probabilities (which are further decomposed into interest-driven and stochastic components). (b) The stochastic exit probabilities are approximated by fitting a Weibull distribution.}
\label{fig:framework}
\end{figure*}

In typical recommender systems, the overall process can generally be divided into three stages: matching, ranking, and re-ranking. The matching stage selects a subset of items relevant to a user’s interests from a large candidate pool. The ranking stage then sorts these items to produce an initial recommendation list. Finally, the re-ranking stage further refines this list by considering richer contextual information or specific system objectives, aiming to improve user satisfaction and overall platform revenue. In this paper, we focus on the re-ranking stage of the recommendation process.

In existing research, mainstream re-ranking methods often adopt the Generator--Evaluator framework. Specifically, the \textbf{Generator} generates multiple possible recommendation lists (i.e., candidate lists) based on the given candidate items, providing a diverse set of choices for subsequent selection. The \textbf{Evaluator} then assesses these lists and selects the one with the highest estimated list-wise value to present to the user.
Formally, let:
\begin{itemize}[leftmargin=*]
    \item $\mathcal{U}$ denote the set of users, with $u \in \mathcal{U}$ denoting the user;
    \item $\mathcal{X}_u$ denote the profile information of user $u$ (e.g., demographics, preferences, historical interactions);
    \item $\mathcal{I}$ denote the set of items, with $i \in \mathcal{I}$ denoting the item;
    \item $\mathcal{X}_i$ denote the profile information of item $i$ (e.g., metadata).
\end{itemize}

For a given user $u$, let $\mathcal{I}_u \subseteq \mathcal{I}$ be the set of candidate item IDs relevant to $u$. Define the enriched candidate item representations as:
\begin{equation*}
\mathcal{C}_u = \{\, \mathcal{X}_i \mid i \in \mathcal{I}_u,\ \,\}.
\end{equation*}
The Generation and Evaluation tasks are defined as follows:

\textbf{Generation Task.} Learn a generator function:
\begin{equation*}
g(\cdot): \big(u,\; \mathcal{X}_u,\; \mathcal{C}_u\big) \rightarrow \{\, l^1_u, l^2_u, \dots, l^k_u \,\},
\end{equation*}
where $k$ is the preset number of generated lists, and each list
\begin{equation*}
l^j_u = [\, i^{j1}_u, i^{j2}_u, \dots, i^{jm}_u \,], 
\quad i^{j*}_u \in \mathcal{I}_u.
\end{equation*}
By producing multiple diverse lists, the Generator enriches the candidate space for the Evaluator.

\textbf{Evaluation Task.} Given the set of generated lists $\{\, l^1_u, l^2_u, \dots, l^k_u \,\}$, where each $l^j_u = [\, i^{j1}_u, i^{j2}_u, \dots, i^{jm}_u \,]$ and their corresponding item side information $\mathcal{X}_{l^j_u} = [\, \mathcal{X}_{i^{j1}_u}, \dots, \mathcal{X}_{i^{jm}_u} ]$, learn an evaluator function:
\begin{equation*}
e(\cdot): \big(u,\; \mathcal{X}_u,\; l_u,\; \mathcal{X}_{l_u} \big) \rightarrow \mathbb{R}.
\end{equation*}
The Evaluator then selects the list with the highest estimated value:
\begin{equation*}
l^*_u = \arg\max_{1 \leq j \leq k} e(u,\; \mathcal{X}_u,\; l^j_u,\; \mathcal{X}_{l^j_u}).
\end{equation*}
Therefore, the Evaluator’s accuracy is crucial for ensuring optimal user experience and achieving business objectives.

%% file: Section/4-Method.tex
\paragraph{Overview}
Traditional evaluators usually estimate the value of the entire list, assuming users will consume all items. However, in reality, users may stop consuming the recommendation list before reaching the end, causing a significant discrepancy between the estimated value and actual consumption values. To address this gap, we propose modeling the user's actual consumption value as a sum of values from sub-lists, weighted by the probability of user exit at each point of the list. Based on this insight, we design the Consumption-Aware List Value Estimation Framework (CAVE), which introduces two key modules: sub-list value estimation and user exit probability estimation.

\subsection{Sub-List Value Estimation}
\input{Section/4-Subsection/1-sub-list}

\subsection{User Exit Probability Estimation}
\input{Section/4-Subsection/2-exit-probability}

\subsection{Training \& Inference Strategy of CAVE}
\input{Section/4-Subsection/3-Training}

%% file: Section/4-Subsection/1-sub-list.tex
As previously discussed, the consumption value of a list can be viewed as the expected value of its sub-lists, weighted by the user exit distribution. To facilitate this, we first estimate the value of each sub-list. Given a user $u$ and a list $l_u$, we represent user features as $\mathcal{X}_u$ and item features as $\mathcal{X}_{l_u}$. To capture the interactions, we merge these features into an embedding sequence $s_u = [e_u^1, e_u^2, \dots, e_u^m]$, where each embedding $e_u^j = (\mathcal{X}_u, \mathcal{X}_{i_u^j})$ corresponds to the $j$-th item in the list.

As shown in Fig.~\ref{fig:framework}(a), considering the sequential dependencies among items within sub-lists, we employ a sequential model $f(\cdot)$ to capture contextual information for each sub-list:
\begin{equation}
[h_u^1, h_u^2, \dots, h_u^m] = f(\mathcal{S}_u),
\label{eq:h}
\end{equation} 
where $h_u^j$ represents the $j$-th sub-list embedding.

We then estimate each sub-list’s value using a scoring head $\phi_s(\cdot)$:
\begin{equation}
r_u^j = \phi_s(h_u^j), \quad j = 1, \dots, m.
\label{eq:sub}
\end{equation}

%% file: Section/4-Subsection/2-exit-probability.tex
Users may exit due to varying interests or random external factors (e.g., fatigue). Therefore, we decompose the user's exit probability into two intuitive components: interest-driven and stochastic exit probabilities.

Interest-driven exit probability represents how likely a user will stop at a certain position based on their interest. Intuitively, if a user's interest at position $j$ decreases, their likelihood of exiting at this position increases. We use the embedding $h_u^j$ obtained from Eq.~\ref{eq:h} to model this with a probability head $\phi_p(\cdot)$:
\begin{equation}
p_u^j = \phi_p(h_u^j), \quad j = 1,\dots, m,
\label{eq:pi}
\end{equation}
where $p_u^j$ represents the user's interest-driven exit probability after item $j$.

On the other hand, the stochastic exit probability reflects user exits driven by external factors unrelated to their interests. To characterize this behavior clearly, we statistically analyze the exist behavior across all users to obtain an empirical exit probability distribution $\bm{P} = [P_1, P_2, \dots, P_k]$. Assuming the interest-driven exit probability is relatively uniform across all users, the empirical exist probability distribution $\bm{P}$ naturally approximates the stochastic exit probability across all users, as demonstrated in Fig.~\ref{fig:framework}(b).

To effectively capture this stochastic pattern without overfitting, we introduce the Weibull distribution~\cite{weibull1951statistical}, which is widely used to model random events and provides intuitive parameters controlling its shape and scale:
\begin{equation}
W(x; \lambda, z) = \frac{z}{\lambda} \left( \frac{x}{\lambda} \right)^{z-1} e^{-(x/\lambda)^z}, \quad x \geq 0,
\end{equation}
where $\lambda$ and $z$ are intuitive parameters controlling scale and shape.

We fit the Weibull parameters intuitively by aligning the empirical discrete exit probability distribution $\bm{P}$ with the continuous Weibull cumulative distribution function (CDF). Specifically, we minimize the intuitive discrepancy between observed and modeled probabilities:
\begin{equation}
\min_{\lambda, z} \sum_{j=1}^{k} \Big( P^j - \big(F(j; \lambda, z) - F(j-1; \lambda, z)\big) \Big)^2,
\end{equation}
where the Weibull CDF intuitively describes the cumulative stochastic exit likelihood:
\begin{equation}
F(x; \lambda, z) = 1 - e^{-(x/\lambda)^z}.
\end{equation}

We then compute the stochastic exit probability at each position $j$ as:
\begin{equation}
p_u^{j,w} = W(j; \lambda, z), \quad j = 1, \dots, m.
\label{eq:pw}
\end{equation}

Finally, to combine the interest-driven and stochastic exit probabilities, we use the power mean:
\begin{equation}
p_u^{j,*} = \sqrt[\alpha]{ \frac{(p_u^{j} + p_u^{j,w})^{\alpha}}{2} }, \quad j = 1, \dots, m,
\label{eq:p*}
\end{equation}
where $\alpha$ controls the combination's smoothness. By adjusting $\alpha$, we can control whether the combined probability $p_u^{j,*}$ leans more toward the smaller or larger value of $(p_u^{j}, p_u^{j,w})$: a smaller $\alpha$ emphasizes lower probabilities, while a larger $\alpha$ places more weight on higher probabilities.

%% file: Section/4-Subsection/3-Training.tex
After estimating each sub-list's value and the corresponding user exit probabilities at each position, we calculate the final consumption value of the generated list $l_u$ as follows:
\begin{equation}
\hat{c}_u = \sum_{j=1}^{\vert s_u \vert} p_u^{j,*} r_u^j.
\end{equation}

\begin{algorithm}[t]
\caption{Training Procedure of CAVE}
\begin{algorithmic}[1]
\Require User set $\mathcal{U}$, generated lists $l_u$, features $\mathcal{X}_u$, $\mathcal{X}_{l_u}$, pre-estimated Weibull parameters $(\lambda, z)$
\For{each user $u$ in $\mathcal{U}$}
\State Construct embedding sequence $s_u$
\State Encode sub-lists: $[h_u^1, \dots, h_u^m]$ by Eq.~~\ref{eq:h}
\State Compute sub-list values: $r_u^j$ by Eq.~~\ref{eq:sub}
\State Compute interest-driven exit probabilities: $p_u^j$ by Eq.~~\ref{eq:pi}
\State Compute stochastic exit probabilities: $p_u^{j,w}$ by Eq.~~\ref{eq:pw}
\State Combine probabilities: $p_u^{j,*}$ by Eq.~\ref{eq:p*}
\State Estimate consumption value $\hat{c}_u = \sum_{j=1}^{m} p_u^{j,*} r_u^j$
\State Compute losses $\mathcal{L}_s$ (Eq.~~\ref{eq:ls}) and $\mathcal{L}_p$ (Eq.~~\ref{eq:lp})
\EndFor
\State Update $\phi_s$ using $\mathcal{L}_s$
\State Freeze $h_u$, $r_u$; update $\phi_p$ using $\mathcal{L}_p$
\end{algorithmic}
\label{alg:train}
\end{algorithm}

To effectively train our model and ensure accurate estimation of the final consumption value, we introduce a training strategy that separately optimizes the scoring and probability modules, each guided by distinct training objectives.

\textbf{Score Head Training}: Since users only provide feedback at their actual exit positions, we train the scoring module $\phi_s(\cdot)$ to directly minimize the difference between the estimated sub-list value at the actual exit position and the user's observed consumption feedback:
\begin{equation}
\mathcal{L}_{s} = \mathbb{E}_{u \in \mathcal{U}}\left[ \left(r_u^m - c_u\right)^2\right],
\label{eq:ls}
\end{equation}
where $c_u$ denotes user $u$'s actual consumption value for the generated list $l_u$.

\textbf{Probability Head Training}: To capture users' interest-driven exit behaviors accurately, the probability module $\phi_p(\cdot)$ is trained by aligning the weighted sum of the estimated sub-list values with the actual observed consumption feedback. This ensures the estimated probabilities realistically reflect the users' exit behaviors:
\begin{equation}
\mathcal{L}_{p} = \mathbb{E}_{u \in \mathcal{U}} \Big[ \Big( \sum_{j=1}^{\vert s_u \vert} (p_u^{j,*} \cdot r_u^j) - c_u \Big)^2 \Big].
\label{eq:lp}
\end{equation}

To mitigate potential conflicts between objectives $\mathcal{L}_{s}$ and $\mathcal{L}_{p}$, we freeze the sub-list embeddings $h_u$ and sub-list values $r_u$ during the training of the probability module $\phi_p(\cdot)$, ensuring stable training.

The comprehensive training process is detailed in Algorithm~\ref{alg:train}, while Algorithm~\ref{alg:infer} illustrates the inference procedure.

\begin{algorithm}[t]
\caption{Inference Procedure of CAVE}
\begin{algorithmic}[1]
\Require Generated list $l_u$, user features $\mathcal{X}_u$, item features $\mathcal{X}_{l_u}$, pre-estimated Weibull parameters $(\lambda, z)$
\State Construct embedding sequence $s_u$ and encode sub-lists $[h_u^1, \dots, h_u^m]$ (Eq.~~\ref{eq:h})
\State Compute sub-list values $r_u^j$ (Eq.~~\ref{eq:sub})
\State Compute interest-driven exit probabilities $p_u^j$ (Eq.~~\ref{eq:pi})
\State Compute stochastic exit probabilities $p_u^{j,w}$ (Eq.~~\ref{eq:pw})
\State Combine probabilities to obtain $p_u^{j,*}$ (Eq.~\ref{eq:p*})
\State Compute final estimated consumption value $\hat{c}_u = \sum_{j=1}^{m} p_u^{j,*} r_u^j$
\State \Return $\hat{c}_u$
\end{algorithmic}
\label{alg:infer}
\end{algorithm}

%% file: Section/5-Benchmark.tex





\input{Section/Table/user_feature}

To support the evaluation of list-wise value estimation and re-ranking strategies, we publicly release three large-scale real-world benchmarks constructed from anonymized short video interaction logs on the Kuaishou platform, spanning April to June 2025.
For each user, we segment their interaction history into multiple independent \textbf{sessions}, where each session corresponds to a sequence of user-item interactions ending with either explicit user exit behavior or inactivity. This ensures that each session reflects authentic user consumption behavior. 

To align with the re-ranking generator, which produces recommendation lists of fixed length, we further divide each session into multiple \textbf{requests}. Each request corresponds to a recommendation list containing up to six items; the final request in a session may contain fewer items if the session length is not divisible by six. This segmentation allows the evaluation framework to effectively model and estimate the consumption value of each generated list.

\subsection{Feature Description}

Each record includes three categories of features:
\input{Section/Table/item_feature}

\textbf{User features} include user ID, gender (0 for unknown, 1/2 for different genders), age level (0--7, each representing a 10-year range from under 10 to over 70), and a user cluster ID (0--26) that captures behavioral grouping.

\textbf{Item features} consist of item ID, video duration (in milliseconds), and four categorical labels from different taxonomy levels, with value ranges 0--232, 0--816, 0--2919, and 0--999, respectively.

\textbf{Request-level features} include session ID, request ID, a list of up to six item IDs, five pre-processed interaction-related feature lists, and three user feedback labels per item: completion rate (0–1), binary positive feedback (e.g., like, follow), and a long-view indicator (whether view time exceeds a threshold).

The structured descriptions of user, item, and request-level features are provided in Tables~\ref{tab:user_feat}, \ref{tab:item_feat}, and \ref{tab:req_feat}, respectively.

\input{Section/Table/session_fino}

These features enable accurate modeling of list-wise consumption behavior and provide rich contextual signals for re-ranking research.

\subsection{Dataset Variants and Sampling Strategies}
We construct three dataset variants using different user sampling strategies to provide comprehensive evaluation scenarios:
\begin{itemize}[leftmargin=*]
\item \textbf{KuaiE-small}: 10K users uniformly sampled for lightweight testing.
\item \textbf{KuaiE-top}: 5K highly active users sampled to represent heavy-user behaviors.
\item \textbf{KuaiE}: 100K users uniformly sampled to reflect typical platform-wide behavior.
\end{itemize}

Each dataset ensures that each user has between 3 and 10 valid sessions, and that each session contains fewer than 5 requests to avoid long-tail cases with minimal interactions (typically only one user action). Detailed statistics are presented in Table~\ref{tab:Kuai_stats}.

\input{Section/Table/datasets}

\input{Section/Table/offline_performance}

%% file: Section/Table/user_feature.tex
\begin{table}[t]
\centering
\caption{User feature description.}
\label{tab:user_feat}
\begin{tabular}{lll}
\toprule
\textbf{Field} & \textbf{Description} & \textbf{Value Range} \\
\midrule
user\_id & User identifier & Integer \\
gender & Gender category & 0: Unknown; 1/2: Gender types \\
age\_level & Age group & 0--7 (per decade) \\
user\_tag & User group type & 0--26 \\
\bottomrule
\end{tabular}
\end{table}

%% file: Section/Table/item_feature.tex
\begin{table}[t]
\centering
\caption{Item feature description.}
\label{tab:item_feat}
\begin{tabular}{lll}
\toprule
\textbf{Field} & \textbf{Description} & \textbf{Value Range} \\
\midrule
item\_id & Item identifier & Integer \\
cat\_1 & Category level 1 & 0--232 \\
cat\_2 & Category level 2 & 0--816 \\
cat\_3 & Category level 3 & 0--2919 \\
cat\_4 & Category level 4 & 0--999 \\
duration & Video length & Milliseconds \\
\bottomrule
\end{tabular}
\end{table}

%% file: Section/Table/session_fino.tex
\begin{table}[t]
\centering
\caption{Request-level features. Each request contains up to 6 items.}
\label{tab:req_feat}
\begin{tabular}{lll}
\toprule
\textbf{Field} & \textbf{Description} & \textbf{Type / Range} \\
\midrule
session\_id & Session identifier & Integer \\
request\_id & Request identifier & Integer \\
item\_id\_list & Items in the request & List of item\_id \\
feat\_list1--5 & Interaction-related features & Float lists \\
label\_completion & Actual completion rate & [0, 1] list \\
label\_positive & Positive feedback & Binary list \\
label\_longview & Long-view indicator & Binary list \\
\bottomrule
\end{tabular}
\end{table}

%% file: Section/Table/datasets.tex
\begin{table}[t]
  \centering
  \caption{Statistics of the released KuaiE benchmark datasets and common datasets.}
  \label{tab:Kuai_stats}
   \resizebox{0.47\textwidth}{!}{
  \begin{tabular}{lrrrr}
    \toprule
    \textbf{Dataset} & \textbf{\#User} & \textbf{\#Item} & \textbf{\#Session} & \textbf{\#Request} \\
    \midrule
    \textbf{KuaiE} (100K)      & 100,000 & 877,455 & 556,235 & 1,634,706 \\
    \textbf{KuaiE-small} (10K) & 10,000  & 294,775 & 42,871  & 140,402   \\
    \textbf{KuaiE-top} (5K)    & 5,000   & 268,520 & 133,887 & 234,555   \\
    \midrule
    \textbf{Arts}    & 9,750   & 1,013,133 & 48,750 & 161,196  \\
    \textbf{Games}    & 30,361   & 2,921,785 & 151,805 & 500,648  \\
    \bottomrule
  \end{tabular}
  }
\end{table}

%% file: Section/Table/offline_performance.tex
\begin{table*}[t]
    \centering
    \caption{Offline Performance Comparison.}
    \resizebox{\textwidth}{!}{

\begin{tabular}{ccrrrrrrr}
    \toprule
\multicolumn{2}{c}{\multirow{2}{*}{\textbf{Dataset}}} & \multicolumn{1}{c}{\multirow{2}{*}{\textbf{DNN}}} & \multicolumn{3}{c}{\textbf{Attention}} & \multicolumn{3}{c}{\textbf{GRN}}\\ 
\cmidrule(lr){4-6} \cmidrule(lr){7-9}
& & & \multicolumn{1}{c}{\textbf{SDN }} & \multicolumn{1}{c}{\textbf{CAVE}} & \multicolumn{1}{c}{Gain} & \multicolumn{1}{c}{\textbf{SDN }} & \multicolumn{1}{c}{\textbf{CAVE }} & \multicolumn{1}{c}{Gain} \\
\midrule
\multirow{3}{*}{\textbf{KuaiE-small}} 
& uAUC (\%) & 60.01 $\pm$ 0.55 & 62.12 $\pm$ 0.38 & \textbf{82.91 $\pm$ 5.75} & 33.47\% $\uparrow$ & 62.74 $\pm$ 0.40 & \textbf{85.82 $\pm$ 5.79} & 36.77\% $\uparrow$ \\
& bAUC (\%) & 60.90 $\pm$ 0.58 & 62.43 $\pm$ 0.08 & \textbf{69.70 $\pm$ 3.08} & 11.64\% $\uparrow$ & 63.05 $\pm$ 0.09 & \textbf{71.53 $\pm$ 3.10} & 13.46\% $\uparrow$ \\
& MSE       & 4.78 $\pm$ 0.09 & 4.57 $\pm$ 0.07 & \textbf{2.42 $\pm$ 0.07} & 46.98\% $\uparrow$ & 4.60 $\pm$ 0.07 & \textbf{2.40 $\pm$ 0.06} & 47.97\% $\uparrow$ \\
\midrule
\multirow{3}{*}{\textbf{KuaiE-top}} 
& uAUC (\%) & 61.16 $\pm$ 0.38 & 67.70 $\pm$ 0.69 & \textbf{86.09 $\pm$ 0.64} & 27.17\% $\uparrow$ & 68.89 $\pm$ 0.68 & \textbf{89.01 $\pm$ 0.67} & 29.18\% $\uparrow$ \\
& bAUC (\%) & 64.54 $\pm$ 0.26 & 67.98 $\pm$ 0.30 & \textbf{72.29 $\pm$ 0.26} & 6.33\% $\uparrow$ & 69.14 $\pm$ 0.30 & \textbf{73.70 $\pm$ 0.27} & 6.59\% $\uparrow$ \\
& MSE       & 4.20 $\pm$ 0.07 & 3.44 $\pm$ 0.06 & \textbf{2.41 $\pm$ 0.06} & 29.95\% $\uparrow$ & 3.46 $\pm$ 0.06 & \textbf{2.39 $\pm$ 0.06} & 31.03\% $\uparrow$ \\
\midrule
\multirow{3}{*}{\textbf{KuaiE}} 
& uAUC (\%) & 59.91 $\pm$ 0.54 & 70.63 $\pm$ 0.54 & \textbf{76.07 $\pm$ 1.78} & 7.70\% $\uparrow$ & 71.01 $\pm$ 0.57 & \textbf{77.85 $\pm$ 1.80} & 9.63\% $\uparrow$ \\
& bAUC (\%) & 60.90 $\pm$ 0.57 & 65.56 $\pm$ 0.33 & \textbf{66.37 $\pm$ 0.78} & 1.23\% $\uparrow$ & 66.03 $\pm$ 0.34 & \textbf{67.39 $\pm$ 0.79} & 2.06\% $\uparrow$ \\
& MSE       & 4.61 $\pm$ 0.07 & 3.82 $\pm$ 0.07 & \textbf{3.25 $\pm$ 0.06} & 15.12\% $\uparrow$ & 3.80 $\pm$ 0.06 & \textbf{3.21 $\pm$ 0.06} & 15.44\% $\uparrow$ \\
\midrule
\multirow{3}{*}{\textbf{Arts}} 
& uAUC (\%) & 70.02 $\pm$ 0.56 & 74.15 $\pm$ 0.42 & \textbf{83.41 $\pm$ 0.60} & 12.50\% $\uparrow$ & 74.88 $\pm$ 0.45 & \textbf{84.55 $\pm$ 0.61} & 12.92\% $\uparrow$ \\
& bAUC (\%) & 70.64 $\pm$ 0.57 & 75.02 $\pm$ 0.40 & \textbf{84.05 $\pm$ 0.59} & 12.04\% $\uparrow$ & 75.73 $\pm$ 0.43 & \textbf{85.12 $\pm$ 0.60} & 12.39\% $\uparrow$ \\
& MSE       & 0.38 $\pm$ 0.02 & 0.32 $\pm$ 0.01 & \textbf{0.24 $\pm$ 0.01} & 25.00\% $\uparrow$ & 0.31 $\pm$ 0.01 & \textbf{0.23 $\pm$ 0.01} & 25.81\% $\uparrow$ \\
\midrule
\multirow{3}{*}{\textbf{Games}} 
& uAUC (\%) & 68.50 $\pm$ 0.61 & 72.02 $\pm$ 0.44 & \textbf{81.30 $\pm$ 0.58} & 12.89\% $\uparrow$ & 72.68 $\pm$ 0.47 & \textbf{82.25 $\pm$ 0.59} & 13.16\% $\uparrow$ \\
& bAUC (\%) & 69.34 $\pm$ 0.59 & 73.10 $\pm$ 0.41 & \textbf{82.04 $\pm$ 0.56} & 12.23\% $\uparrow$ & 73.76 $\pm$ 0.43 & \textbf{83.18 $\pm$ 0.57} & 12.78\% $\uparrow$ \\
& MSE       & 0.39 $\pm$ 0.02 & 0.34 $\pm$ 0.01 & \textbf{0.27 $\pm$ 0.01} & 20.59\% $\uparrow$ & 0.33 $\pm$ 0.01 & \textbf{0.26 $\pm$ 0.01} & 21.21\% $\uparrow$ \\
\bottomrule

\end{tabular}
    }
\label{tab:offline}%
\end{table*}

%% file: Section/6-Experiment.tex
In this section, we conduct both offline and online experiments to validate the effectiveness of CAVE. For the offline evaluation, we compare CAVE against state-of-the-art estimation methods on three released KuaiE benchmarks as well as two public datasets from Amazon. For the online evaluation, we deploy CAVE in a real-world industrial video recommendation platform and perform an online A/B test.

\subsection{Offline Experiments}\label{sec: experiments_offline}
\input{Section/6-Subsection/1-offline}

\subsection{Live Experiments}\label{sec: experiments_live}
\input{Section/6-Subsection/2-online}

%% file: Section/6-Subsection/1-offline.tex
\subsubsection{Datasets and Offline Experiments Setting}
We use five datasets in our offline experiments: three released KuaiE benchmarks described in Section~\ref{sec:bench}, and two widely adopted datasets from Amazon---Arts and Games~\cite{ni2019justifying, zhang2023robust}. For the Amazon datasets, we treat the list of user reviews as the exposure list and the corresponding ratings as feedback signals. Based on interaction timestamps, we segment each user's interaction history into multiple sessions. To maintain consistency with the KuaiE datasets, we further divide each session into multiple requests of length 6. We filter out users with fewer than three sessions. For all five datasets, we use the last session of each user as the test set, the second-to-last as the validation set, and the remaining sessions as the training set. Detailed statistics for all datasets are summarized in Table~\ref{tab:Kuai_stats}.

\input{Section/Table/offline_rate}

\subsubsection{Evaluation Protocol}
To evaluate the effectiveness of CAVE, we adopt three metrics: user-wise AUC (uAUC), batch-wise AUC (bAUC), and mean squared error (MSE). \textbf{uAUC} measures the generalized AUC across all candidate lists for each individual user. It captures the model’s ability to rank lists correctly within each user. \textbf{bAUC} computes the generalized AUC over all samples within a test batch. This reflects the global ranking performance across the test set. \textbf{MSE} evaluates the regression accuracy by computing the mean squared error between the predicted list value and the ground-truth consumption feedback. It directly reflects the fidelity of the model’s estimation.

\subsubsection{Baselines}
We compare our proposed framework with the following baselines to evaluate the effectiveness of CAVE:
\begin{itemize}[leftmargin=*]
    \item \textbf{DNN}: A point-wise estimation model that independently predicts the value of each item in the list based solely on user and item features, which ignores the contextual dependencies.
    \item \textbf{SDN}: A list-wise estimation baseline that incorporates contextual information across the list.
    \item \textbf{CAVE}: Our proposed consumption-aware framework that models both sub-list values and user exit probabilities to estimate the actual consumption value of a list.
\end{itemize}
To assess the impact of different sequential modeling strategies, we implement two variants of both SDN and CAVE by adopting (1)~an \textbf{Attention-based encoder}~\cite{vaswani2017attention} and (2)~a \textbf{Generative Rerank Network (GRN)-based encoder}~\cite{feng2021grn} as their sequential modeling components. This results in five methods in total: DNN, SDN-Attention, SDN-GRN, CAVE-Attention, and CAVE-GRN. Our implementation is available at the following link\footnote{\url{https://github.com/Kaike-Zhang/CAVE}}.

\subsubsection{Main Results}\label{sec: experiments_offline_main_result}
Table~\ref{tab:offline} reports the offline performance comparison on five datasets under the completion rate label, using three metrics: uAUC, bAUC, and MSE. We observe that \textbf{CAVE consistently outperforms all baselines}. On the three KuaiE datasets, CAVE achieves an average improvement of \textbf{25.4\%} in uAUC, \textbf{10.6\%} in bAUC, and a \textbf{31.7\%} reduction in MSE compared to the best-performing baseline. On the two Amazon datasets, CAVE also demonstrates strong performance, with average gains of \textbf{13.0\%} in uAUC, \textbf{12.6\%} in bAUC, and a \textbf{23.2\%} reduction in MSE.

To further verify the robustness of our approach across different value labels, we evaluate all methods on the \textit{positive feedback} label, which indicates whether the user performed a positive action (e.g., like, follow) on an item. \textbf{We observe that users are more likely to give positive feedback earlier in the session, which aligns well with the assumptions of the Weibull distribution used in our exit modeling. This synergy allows CAVE to better capture user behavior dynamics, leading to significant gains.} As shown in Table~\ref{tab:offline-positive}, CAVE again significantly outperforms the baselines. On KuaiE-small, CAVE improves uAUC by \textbf{62.4\%}, bAUC by \textbf{23.6\%}, and reduces MSE by \textbf{34.2\%} compared to the best baseline. On KuaiE-top, the improvements are also significant: \textbf{35.9\%} in uAUC, \textbf{18.1\%} in bAUC, and \textbf{42.5\%} reduction in MSE.

These results validate two key conclusions. First, list-wise modeling that captures contextual dependencies (e.g., SDN and CAVE) significantly outperforms point-wise baselines (e.g., DNN), highlighting the importance of sequential modeling in list evaluation. Second, by explicitly incorporating user exit behavior, CAVE achieves more accurate estimation of actual consumption value, leading to more reliable list selection. The consistent gains across both value labels (completion rate and positive feedback) further demonstrate the generalizability and effectiveness of the proposed consumption-aware framework.

\subsubsection{Ablation Study} \label{sec: experiments_ablation}
To better understand the contribution of each component in CAVE, we conduct an ablation study on the three KuaiE datasets. As shown in Table~\ref{tab:ablation}, we consider the following two ablated variants:
\begin{itemize}[leftmargin=*]
\item \textbf{w/o-wb}: Removes the stochastic exit component modeled by the Weibull distribution.
\item \textbf{w/o-intr}: Removes the interest-driven exit component and retains only the stochastic part.
\end{itemize}
We observe that removing either component results in a clear performance drop across all datasets and both metrics (uAUC and bAUC). Specifically, compared to the full CAVE model, removing the stochastic modeling (w/o-wb) leads to an average uAUC drop of 6.2\%, while removing the interest-driven component (w/o-intr) causes an even larger degradation, especially on the KuaiE dataset (uAUC drops from 76.07\% to 66.10\%).
These results confirm that both components---interest-driven and stochastic---are essential for accurately modeling user exit behavior. Their combination enables CAVE to provide more reliable consumption value estimation and improved re-ranking performance.

\subsubsection{Case Study}
\begin{figure}[t]
\centering
\includegraphics[width=\linewidth]{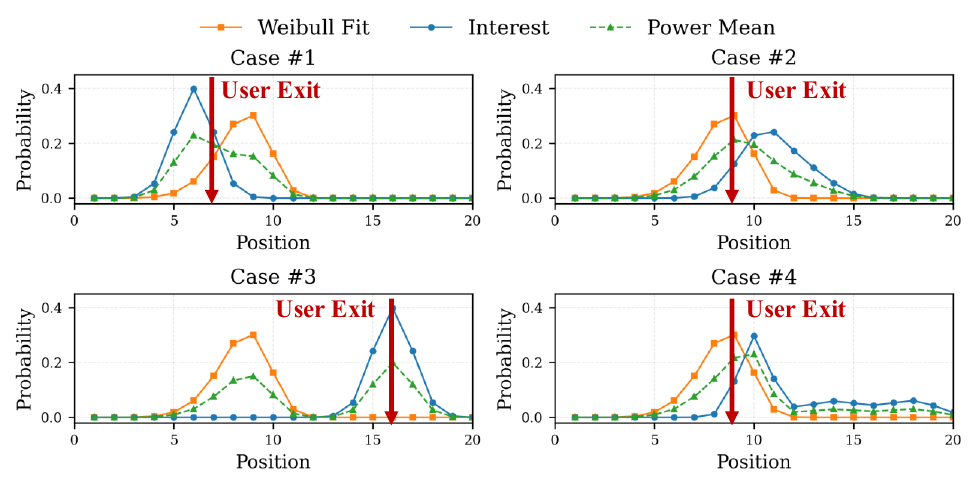}
\caption{Case study of user exit probability estimation.}\label{fig: user_case}
\end{figure}

Figure.~\ref{fig: user_case} presents four representative user cases to visualize the estimated interest-driven exit probability, stochastic exit probability (modeled by the Weibull distribution), and their final combination with $\alpha=2$. The actual user exit positions are also marked for reference. We observe that the combined exit probabilities predicted by CAVE consistently peak around the user's actual exit position across different cases. This demonstrates the effectiveness of our probabilistic modeling: by integrating both personalized interest and stochastic uncertainty, CAVE can accurately capture real user exit behaviors.

\input{Section/Table/ablation}

%% file: Section/Table/offline_rate.tex
\begin{table}[t]
    \centering
    \caption{Offline Performance Comparison on label:positive.}
    \resizebox{0.47\textwidth}{!}{
\begin{tabular}{crrrr}
    \toprule
    \textbf{Metric} & \textbf{DNN} & \textbf{SDN} & \textbf{CAVE} & \textbf{Gain} \\
    \midrule
    \multicolumn{5}{c}{KuaiE-small} \\
    \midrule
    uAUC (\%) & 57.42 $\pm$ 0.48 & 59.18 $\pm$ 0.42 & 96.09 $\pm$ 0.08 & 62.39\% $\uparrow$ \\
    bAUC (\%) & 59.62 $\pm$ 0.44 & 61.20 $\pm$ 0.40 & 75.63 $\pm$ 0.43 & 23.57\% $\uparrow$ \\
    MSE       & 6.89 $\pm$ 0.15  & 6.52 $\pm$ 0.16  & 4.29 $\pm$ 0.13  & 34.22\% $\downarrow$ \\
    \midrule
    \multicolumn{5}{c}{KuaiE-top} \\
    \midrule
    uAUC (\%) & 70.42 $\pm$ 0.42 & 72.08 $\pm$ 0.41 & 97.94 $\pm$ 0.45 & 35.89\% $\uparrow$ \\
    bAUC (\%) & 64.86 $\pm$ 0.39 & 65.70 $\pm$ 0.38 & 77.61 $\pm$ 1.36 & 18.11\% $\uparrow$ \\
    MSE       & 6.85 $\pm$ 0.18  & 7.02 $\pm$ 0.17  & 4.04 $\pm$ 0.25  & 42.45\% $\downarrow$ \\
    \bottomrule
\end{tabular}
    }
\label{tab:offline-positive}
\end{table}

%% file: Section/Table/ablation.tex
\begin{table}[t]
\centering
\caption{Ablation study of CAVE across datasets.}
 \resizebox{0.47\textwidth}{!}{
\begin{tabular}{lcccccc}
\toprule
\multirow{2}{*}{Model} & \multicolumn{2}{c}{KuaiE-small (\%)} & \multicolumn{2}{c}{KuaiE-top (\%)}  & \multicolumn{2}{c}{KuaiE (\%)} \\
\cmidrule(lr){2-3} \cmidrule(lr){4-5} \cmidrule(lr){6-7}
& uAUC  & bAUC & uAUC & bAUC & uAUC & bAUC \\
\midrule
\textbf{CAVE} & \textbf{82.91} & \textbf{69.70} & \textbf{86.09} & \textbf{72.29} & \textbf{76.07} & \textbf{66.37} \\
w/o-wb & 75.30 & 64.80 & 78.80 & 66.90 & 72.30 & 64.30 \\
w/o-intr & 77.00 & 67.20 & 80.90 & 69.10 & 66.10 & 59.60 \\
\bottomrule
\end{tabular}
}
\label{tab:ablation}
\end{table}

%% file: Section/6-Subsection/2-online.tex
\subsubsection{Experimental Settings}

\begin{figure}[t]
\centering
\includegraphics[width=\linewidth]{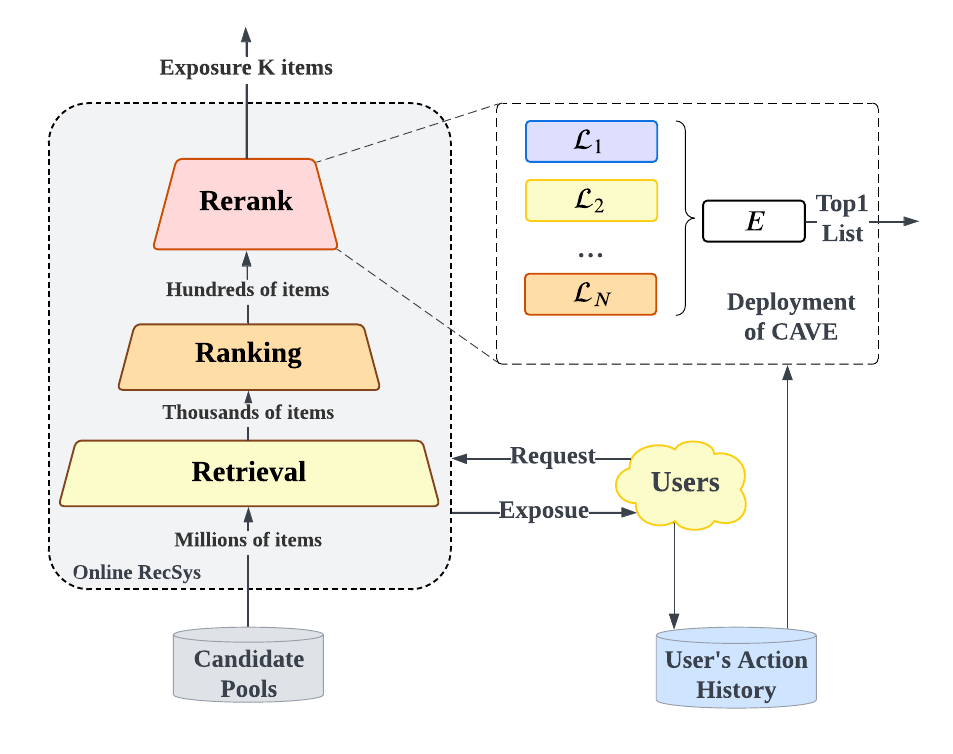}
\caption{The entire workflow consists of a relevant video retrieval stage, an efficient initial ranking stage, a refined ranking stage, and a final reranking . The CAVE is deployed in the Evaluator of the reranking stage.}\label{fig: online_workflow}
\end{figure}

\begin{table*}[t]
\caption{Online performances of CAVE and all results are statistically significant.}
\centering
\begin{tabular}{ccccccc}
\toprule
\textbf{Method}&   Watch Time & Effective View & Follow & Collect  & LT7 & LT30\\
\midrule

\textbf{CAVE} \textit{v.s.} \textbf{baseline} & +0.068$\%$ & +0.328 $\%$ & +0.319$\%$ & +0.328$\%$ & +0.065$\%$ & +0.068$\%$ \\
\bottomrule
\end{tabular}
\label{tab: online_results}
\end{table*}
We conduct an online A/B test on a large scale real-world industrial video recommendation platform Kuaishou to evaluate the effectiveness of CAVE.
The platform serves videos for over half a billion users daily, with an item pool comprising tens of millions of videos.
Figure~\ref{fig: online_workflow} illustrates the detailed implementation of our online recommendation workflow. Specifically, we are considering the recommendation task that recommends a list of 6 items for each user request, which corresponds to the reranking stage. During the reranking stage, our baseline follows the Multi-Generator Evaluator(MG-E)~\cite{yang2025comprehensive} framework, where Generator produces several candidate lists, and the Evaluator selects the top-ranked list for exposure. The input candidate set size at this stage is approximately 120 items~(filtered by previous retrieval and ranking stages), and the final output list size is $K=6$. Our proposed approach CAVE is deployed within the Evaluator module to determine the optimal list to serve.

\begin{figure}[t]
\centering
\includegraphics[width=\linewidth]{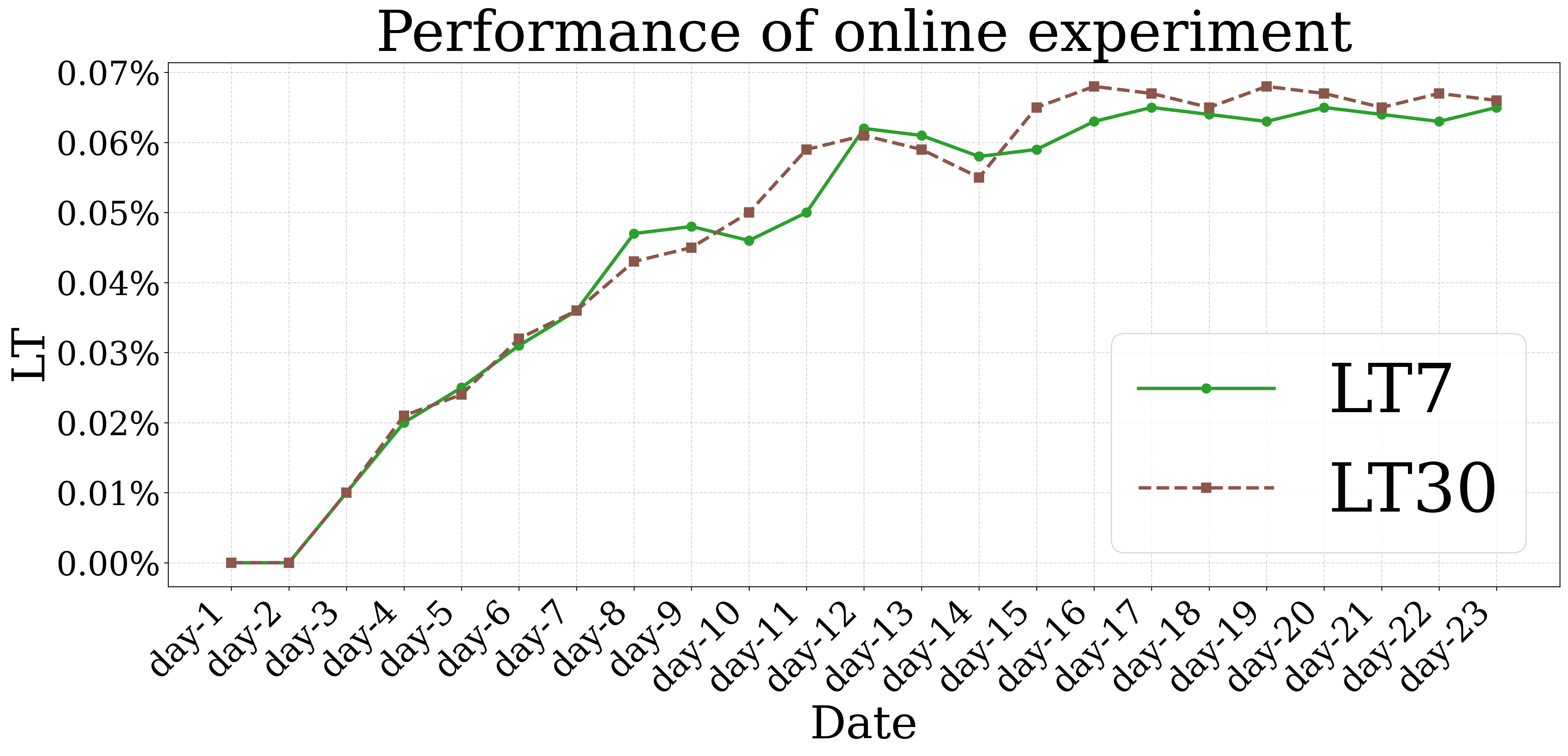}
\caption{Online A/B test performance of Enter LT }\label{fig: LT}
\end{figure}

\subsubsection{Evaluation Protocol}

For each online experiment, we randomly separate the data traffic into eight buckets, each accounting for relatively 1/8 of the total traffic~(each bucket over tens of millions of users) are compared as the baseline MG-SDN and our proposed MG-CAVE.  In the training procedure, we compare the uAUC and bAUC about reward in the key components with empirical weight, including the average watch time, effective/long/finish view count, effective interaction~(including like, follow, share). In the online AB test phase, each experiment is subjected at least 14 days to ensure the reliability. We leverage the common metrics to evaluate the online performance. Note that the ``LT7'' means 7-day Lifetime, which is a key metric that indicates long-term daily active users~(DAU) and user retention benefits online in the next week, ``watch time'' metric expresses the average length the user watches a video without discretization, ``Effective View'' metric represent the total account of views, and all other metrics express the rate of a certain behavior on recommended videos.

\begin{table}[t]
\caption{Performance of Online Training on Data Streams with Convergent Results}
\centering
\begin{tabular}{cccccc}
\toprule
\textbf{Method} &   uAUC (\%) &  Gain & bAUC (\%) &  Gain\\
\midrule

\textbf{SDN} & 67.75 & -- & 67.53 & -- \\
\textbf{CAVE} & 69.05 & 1.92 $\%$ & 69.43 & 2.81 $\%$ \\
\bottomrule
\end{tabular}
\label{tab: online_results}
\end{table}

\subsubsection{Results Analysis}

We keep all experiment online for two weeks and summarize the results in Table \ref{tab: online_results} , which shows that CAVE outperforms the baseline significantly in all metrics and our model achieves significant improvements of \textbf{0.065$\%$ / 0.068$\%$} in LT7/LT30 respectively, proving the stable and consistent improvement on user satisfaction in the long run. Furthermore, we tested the application of CAVE in estimating interactions in the online environment and ensembled the results into the existing score. This directly led to a \textbf{7.38$\%$} increase in the online metric of at least one interaction, which further reinforces our ideas in Section \ref{sec: experiments_offline_main_result}.
\textbf{CAVE is now deployed online with full users at Kuaishou, serving all of traffic every day.}

%% file: Section/7-Conclusion.tex
In this work, we propose \textbf{CAVE}, a novel Consumption-Aware list Value Estimation framework for re-ranking in recommender systems. Unlike traditional evaluators that estimate the value of entire generated lists without accounting for user behavior, CAVE explicitly models user exit probabilities to bridge the gap between generation value and actual consumption value. By decomposing exit probability into an interest-driven component and a stochastic component (modeled via the Weibull distribution), and by jointly estimating sub-list values and exit likelihoods, CAVE provides a more accurate estimation of list consumption value.
To support research in this direction, we also release three large-scale, real-world benchmarks collected from the Kuaishou platform, featuring millions of list-wise exposures with rich user/item features and feedback labels. Extensive experiments on both offline and online settings demonstrate that CAVE consistently outperforms strong baselines, highlighting the importance of incorporating user exit behavior in list evaluation.